\documentclass[11pt,a4paper]{article}
\usepackage{graphicx}
\usepackage{txfonts}
\usepackage{wasysym}

\title{Lane--Emden stars, selfgravitating disks and the Sobolev inequality  }

\date{}

\author{Patryk Mach and Edward Malec \vspace{1ex}\\
\normalsize\textit{M. Smoluchowski Institute of Physics, Jagiellonian University},\\
\normalsize\textit{Reymonta 4, 30-059 Krak\'ow, Poland}}

\begin{document}

\maketitle

\begin{abstract}
 We estimate  the minimal mass of selfgravitating polytropic disks using the famous Sobolev inequality. This bound resembles the well known mass formula for Lane--Emden stars. For ideal gas with the polytropic index $n = 3$ the minimal mass is not smaller than the Jeans mass. The accuracy of the estimate is verified in a number of numerical examples. The bound works well for heavy selfgravitating disks and is less useful for light disks.
\end{abstract}

\section{Introduction}

Models of static spherically symmetric configurations of selfgravitating polytropic fluids   can be reduced to the analysis of the so-called Lane--Emden equations. They fall into the class of nonlinear ordinary differential equations. Their solutions can be obtained by means of simple numerical methods or  even analytically for some polytropic exponents.

In contrast to that, equations describing axially symmetric models of selfgravitating polytropic fluids undergoing a stationary rotation are in general intractable analytically. They can be written in the form of the Poisson equation for the gravitational potential and equations of hydrodynamics that can be formally integrated to yield an algebraic relation between the specific enthalpy, angular velocity, and the gravitational potential. Another possibility, less useful computationally, but revealing the mathematical structure of the problem, is to express them as a single nonlinear elliptic equation for the specific enthalpy (or the mass density) with Dirichlet conditions imposed on the a priori unknown boundary of the disk.

We derive  analytical results on accretion disks using   simple physical information and certain functional analytic methods. There exists the so-called Sobolev inequality \cite{rosen_71}. In this paper we demonstrate that it can be used in order to estimate the mass of selfgravitating toroids. This approach  works for those rotation laws that possess a centrifugal potential $\Phi_\mathrm{c}$ such that $\Delta \Phi_\mathrm{c} \le 0$  (here $\Delta $ is the Laplacian). The equation of state of fluid is $p = K \rho^{1 + 1/n}$, where $p$ is the pressure, $\rho$ the density, and $n \geq 3$. In analogy to the well known Lane-Emden case, the estimate requires the knowledge of the maximal  density and  temperature of the configuration. The same bound is obtained for rotating polytropic stars. It yields rigorously, for ideal gas with the polytropic index $n= 3$, that the mass of  stationary systems is not smaller than the Jeans mass. That is probably the first rigorous derivation of the Jeans inequality.

The order of the paper is as follows. In the next section all relevant equations are displayed. Section 3 gives the estimate for the mass of a selfgravitating disk. In section 4 we discuss the spherically symmetric case. There emerges a striking   similarity between the expression for the mass of a Lane--Emden star and the estimate that is proved in Section 3. Section 5 compares results of the numerical solutions describing axially symmetric disks and the analytical estimate of this paper. This comparison shows that our approach is particularly robust in the strongly nonlinear regime, when the disk masses are much larger than the central mass. Final Section contains a brief summary.

\section{Notation and equations}

Consider a disk of perfect fluid rotating around a central point mass. In this case stationary Euler equations can be written as
\begin{equation}
\label{euler_eq}
(\mathbf U \cdot \nabla) \mathbf U = - \nabla \Phi - \frac{1}{\rho} \nabla p.
\end{equation}
Here $\mathbf U$ denotes the fluid velocity; $\Phi$ is the gravitational potential; $\rho$ denotes the density and $p$ the pressure of the fluid. For a selfgravitating disk the gravitational potential satisfies
\begin{equation}
\label{poisson_eq}
\Delta \Phi = 4 \pi G \rho,
\end{equation}
where $G$ is a gravitational constant.

Let $(r, \phi, z)$ denote cylindrical coordinates. We will consider purely rotating, axially symmetric configurations, so that $\mathbf U = \omega \partial_\phi$, where $\omega$ is the angular velocity.

In order to find solutions of the above system of equations it is customary to assume a fixed form of the rotation law $\omega = \omega(r)$ and the equation of state $p = p(r)$. There is a vast literature on the numerical solutions of such problem (see e.g.~\cite{eriguchi_muller_85}). In this paper we are interested in establishing general analytic bounds on the mass of the disk.

In what follows we will specialize to the polytropic equation of state $p = K \rho^{1 + 1/n}$, where $K$ and $n$ are constant. Let us introduce the specific enthalpy $h$, so that $dh = dp /\rho$. For the polytropic equation of state $h = K (1 + n) \rho^{1/n}$. 

Computing the divergence of Eq.~(\ref{euler_eq}) yields
\begin{equation}
\label{h_eq}
\Delta h = - 4 \pi G \rho + \frac{1}{r} \partial_r \left( r^2 \omega^2 \right) = - C h^n - \Delta \Phi_\mathrm{c},
\end{equation}
where we have introduced the centrifugal potential
\begin{equation}
\label{centrifugal_pot}
\Phi_\mathrm{c} = - \int^r dr^\prime r^\prime \omega^2(r^\prime),
\end{equation}
and a constant $C = 4 \pi G / (K (1 + n))^n$.

Notice that Eq.~(\ref{h_eq}) can be further simplified by assuming the so-called $v$-const rotation law, i.e, $\omega = v_0/r$, where $v_0$ is a constant. In this case the centrifugal term on the right-hand side vanishes, and we have
\begin{equation}
\label{h_eq_simple}
\Delta h = - C h^n.
\end{equation}
This simple form does not imply that the rotation does not influence the structure of the disk. It only means that $  \Phi_\mathrm{c}$ is a harmonic function inside the volume occupied by a disk. Let us point out that Eq.~(\ref{h_eq_simple}) is still difficult to solve, because the boundary condition $h = 0$ is to be posed on an unknown boundary of the disk.

In this paper we will reserve the symbol $\Omega$ for the domain in $\mathbb R^3$ occupied by the disk. The disk boundary will be denoted by $\partial \Omega$.

Eq.~(\ref{h_eq_simple}) is also valid for a static polytropic star. In that case, if we assume isotropy, introduce spherical coordinates and properly rescale variables, it reduces to the well known Lane--Emden equation.

\section{Estimates of the disk mass}
 
Assume that $\Delta \Phi_\mathrm{c} \leq 0$. For the class of rotation laws of the form $\omega = \mathrm{const}/r^p$ this implies that $p \leq 1$, so that the $v$-const rotation is a limiting case.

Let us multiply both sides of Eq.~(\ref{h_eq}) by $h$ and integrate over $\Omega$. It is easy to observe that
\begin{eqnarray}
- \int_\Omega d^3 x h \Delta h & = & \int_\Omega d^3 x |\nabla h|^2 = C \int_\Omega d^3 x h^{n+1} + \int_\Omega d^3 x h \Delta \Phi_\mathrm{c} \nonumber \\
& \leq & C  \int_\Omega d^3 x h^{n+1},
\label{h_integrals}
\end{eqnarray}
where the left-hand side has been integrated by parts, and we have used the fact that $h = 0$ on $\partial \Omega$. Further steps are adapted from~\cite{malec_88}. The last integral in Eq.~(\ref{h_integrals}) can be estimated making use of the H\"{o}lder inequality. For $n > 1$ we have
\[ \int_\Omega d^3x h^{n+1} = \int_\Omega d^3 x h^{n-1} h^2 \le \left\Vert h^{n-1} \right\Vert_{L^{3/2}(\Omega)} \left\Vert h^2 \right\Vert_{L^3(\Omega)}. \]
Finally with the help of the Sobolev inequality
\[ \left\Vert h \right\Vert_{L^6(\Omega)} \le C(3,2) \left\Vert \nabla h \right\Vert_{L^2(\Omega)} \]
we arrive at
\[ \left\Vert \nabla h \right\Vert^2_{L^2(\Omega)} \le C C^2(3,2) \left\Vert h^{n-1} \right\Vert_{L^{3/2}(\Omega)} \left\Vert \nabla h \right\Vert^2_{L^2(\Omega)}. \]
Thus
\[ \left\Vert h^{n-1} \right\Vert_{L^{3/2}(\Omega)} = \left( \int_\Omega d^3 x h^\frac{3(n-1)}{2} \right)^\frac{2}{3} \ge \frac{1}{C C^2(3,2)}. \]
The Sobolev constant $C(3,2) = 4^{1/3} / \left( \sqrt{3} \pi^{2/3} \right)$ is a universal number in $\mathbb R^3$ (cf.~\cite{rosen_71, talenti_76}). The specific enthalpy $h$ can be extended to a function defined on $\mathbb R^3$ by setting $h = 0$ outside $\Omega$. Such an extension belongs to $W_0^{1,2}(\mathbb R^3)$, i.e., the closure of the set of compactly-supported $C^\infty(\mathbb R^3)$ functions in the Sobolev space $W^{1,2}(\mathbb R^3)$.

The mass of a disk is given by
\[ M = \int_\Omega d^3 x \rho = \frac{C}{4 \pi G} \int_\Omega d^3 x h^n. \]

Let $I$ be
\[ I = \int_\Omega d^3 x h^\frac{3(n-1)}{2} = \int_\Omega d^3 x h^\frac{n-3}{2} h^n. \]
For $n > 3$ the value of $I$ can be estimated as
\[ I < h_\mathrm{max}^\frac{n-3}{2} \int_\Omega d^3x h^n, \]
where $h_\mathrm{max}$ denotes the maximum value of the enthalpy within the disk. The reversed inequality holds for $n < 3$.

A combination of the above results gives the lower bound on the mass of the disk in the form
\begin{equation}
\label{mass_ineq}
M > \left( 4 \pi G \sqrt{C} C^3(3,2) h_\mathrm{max}^\frac{n - 3}{2} \right)^{-1} = \left( \frac{K (1 + n)}{4 \pi G} \right)^\frac{3}{2} \rho_\mathrm{max}^{- \frac{n - 3}{2n}} C^{-3}(3,2)
\end{equation}
valid for $n > 3$ and $\Delta \Phi_\mathrm{c} \leq 0$. Here, similarly, $\rho_\mathrm{max}$ denotes the maximum of the density within the disk.

The obtained result can be also understood as a bound for the maximum temperature in the gas configuration. For the ideal gas $T = p \mu m_\mathrm{p} / \left( \rho k_\mathrm{B} \right)$, where $\mu$ is the mean molecular weight, $m_\mathrm{p}$ denotes the mass of the proton, and $k_\mathrm{B}$ is the Boltzmann constant. Inequality (\ref{mass_ineq}) can be now written as
\[ T_\mathrm{max} < 4 \pi G \mu m_\mathrm{p} M^\frac{2}{3} \rho_\mathrm{max}^\frac{1}{3} C^2(3,2) / \left((1 + n) k_\mathrm{B} \right), \]
where $T_\mathrm{max}$ denotes the maximal temperature of gas.

Yet another consequence of (\ref{mass_ineq}) can be obtained for the value $n=3$ of the polytropic index. Let $\bar \rho $ and $\bar T$ denote volume averaged mass density and temperature, respectively.
For the ideal gas one gets, after simple calculations involving a H\"older inequality:
\begin{equation}
\label{holder}
\bar T \le K{\mu m_\mathrm{p} \over k_B} {\bar \rho}^\frac{1}{n}.
\end{equation} 
Inserting that into (\ref{mass_ineq}) yields the estimate
\begin{equation}
\label{Jeans}
M > \left( \frac{k_B}{\pi G \mu m_\mathrm{p}} \right)^\frac{3}{2}  \frac{\bar T^\frac{3}{2}}{\sqrt{\bar \rho}} C^{-3}(3,2),
\end{equation}
or, writing $C^{-3}(3,2)$ explicitely,
\begin{equation}
\label{Jeans1}
M > \frac{3 \sqrt{3 \pi}}{4} \left( \frac{k_B}{G \mu m_\mathrm{p}} \right)^\frac{3}{2}  \frac{\bar T^\frac{3}{2}}{\sqrt{\bar \rho}} \equiv M_\mathrm{S}.
\end{equation}
The Jeans mass is usually expressed as 
\begin{equation}
\label{Jeans2}
M_\mathrm{J} = \beta \left( \frac{k_B}{G \mu m_\mathrm{p}} \right)^\frac{3}{2} \frac{\bar T^\frac{3}{2}}{\sqrt{\bar \rho}},
\end{equation}
where $\beta $ is a constant (dependent on the convention of the Jeans mass) of the order of 1/2.
Bound systems should possess a mass $M > M_\mathrm{J}$, according to a derivation that traditionally suffers from severe gaps. Notice that the Jeans mass $M_\mathrm{J}$ is close to $M_\mathrm{S}$, the right hand side of (\ref{Jeans1}); indeed, $M_\mathrm{S} = M_\mathrm{J} \times  3 \sqrt{3 \pi} / (4 \beta)$. Therefore for bound systems we obtain rigorously $M > 3 \sqrt{3 \pi} / (4 \beta) M_\mathrm{J}$. 
 
\section{Lane--Emden stars}

\begin{figure}[t!]
\begin{center}
\includegraphics[width=10cm]{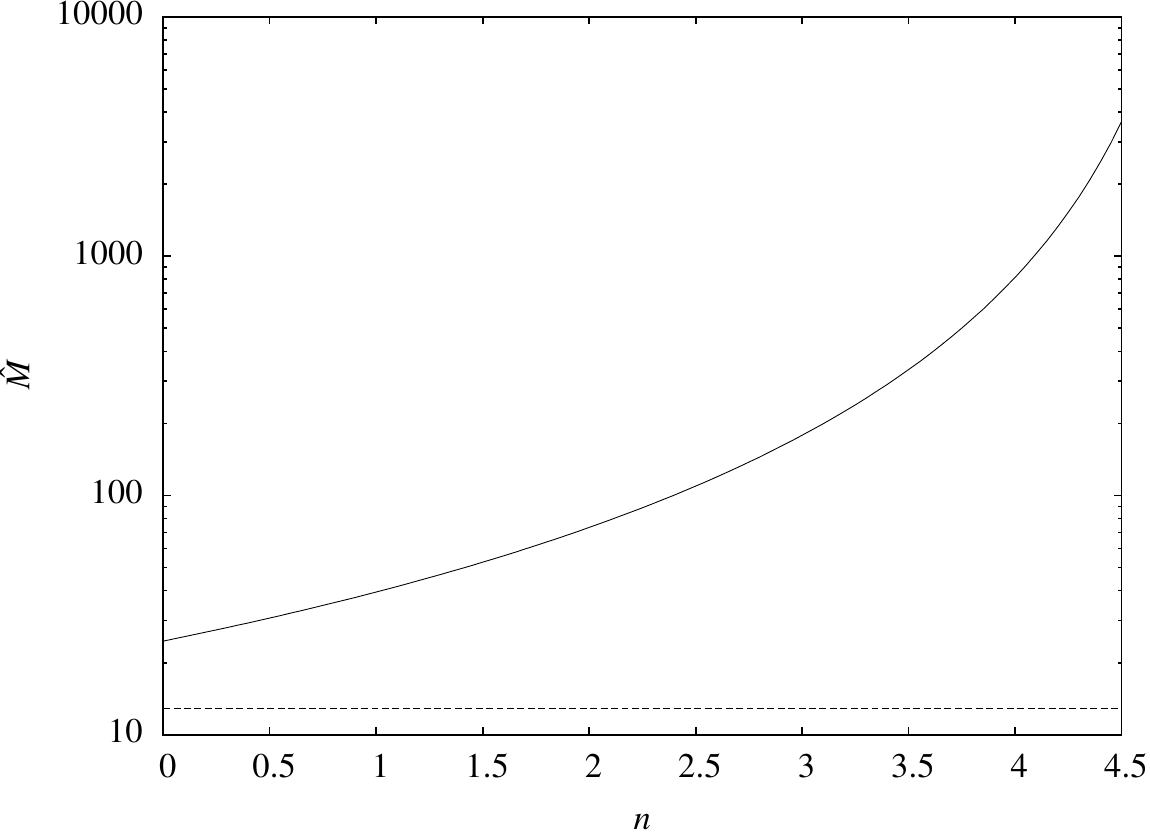}
\end{center}
\caption{The integral $\hat M$ as a function of $n$. The horizontal line represents the value of $C^{-3}(3,2) = 12.821$.}
\label{fig_lane_mass}
\end{figure}

Inequality (\ref{mass_ineq}) is obviously satisfied for static polytropic stars, described entirely in terms of the so-called Lane--Emden functions. 

Assuming the spherical symmetry Eq.~(\ref{h_eq}) can be reduced to the Lane--Emden equation
\begin{equation}
\label{lane_emden_eq}
\frac{1}{\xi^2} \frac{d}{d\xi} \left( \xi^2 \frac{d \theta}{d \xi} \right) + \theta^n = 0,
\end{equation}
where
\[ \xi = \sqrt{\frac{4 \pi G \rho_\mathrm{max}^{1 - 1/n}}{K(1 + n)}} r, \]
$\rho = \rho_\mathrm{max} \theta^n$, and $r$ denotes the distance from the center of the star. Since $\rho_\mathrm{max} = \rho(r = 0)$, we have $\theta(\xi = 0) = 1$. A textbook exposition of the theory of Lane--Emden equation can be found in~\cite{chandrasekhar}.

Let the first zero of $\theta$ (if present) be denoted by $\xi_0$. If the Lane--Emden function corresponding to a given index $n$ has no zeros, we assume $\xi_0 = \infty$. A radius corresponding to $\xi_0$ will be denoted by $R$. The mass of the star can be computed as
\begin{equation}
\label{lane_emden_mass}
M = \int_0^R 4 \pi r^2 \rho dr = \left( \frac{K (1 + n)}{4 \pi G} \right)^\frac{3}{2} \rho_\mathrm{max}^{- \frac{n - 3}{2n}}  \hat M ,
\end{equation}
where
\[ \hat M =  \int_0^{\xi_0} 4 \pi \xi^2 \theta^n d \xi = - 4 \pi \left( \xi^2 \frac{d \theta}{d \xi} \right)_{\xi_0}. \]
Since analytical expressions for $\theta$ are only known for $n = 0$, 1 and 5, the values of $\hat M$ have to be computed numerically.

The similarity between Eqs.~(\ref{mass_ineq}) and (\ref{lane_emden_mass}) is remarkable. We see that for $n > 3$ there must be $\hat M >  C^{-3}\left( 3,2\right)$.

The general bound on $\hat M$ can be obtained by observing that $\hat M$ attains its minimum value for $n = 0$. In this case the solution $\theta$ is known analytically and $\hat M = \hat M_\mathrm{min} = 16 \sqrt{6} \pi / 5 = 24.625 $. Numerical values of $\hat M$ for different indices $n$ are shown on Fig.~\ref{fig_lane_mass}. The factor $\hat M$ is close to $C^{-3}(3,2)$ for $n$ close to zero, which is outside the validity zone of our estimate. The solutions of the Lane--Emden equations are, however, relatively easy to be obtained numerically, and this case can only be treated as a demonstration of the validity of Eq.~(\ref{mass_ineq}). In the next section we will present some numerical results for the selfgravitating disks.

\section{Selfgravitating disks}

\begin{figure}[t!]
\vspace{-1cm}
\begin{center}
\includegraphics[width=10cm,angle=-90]{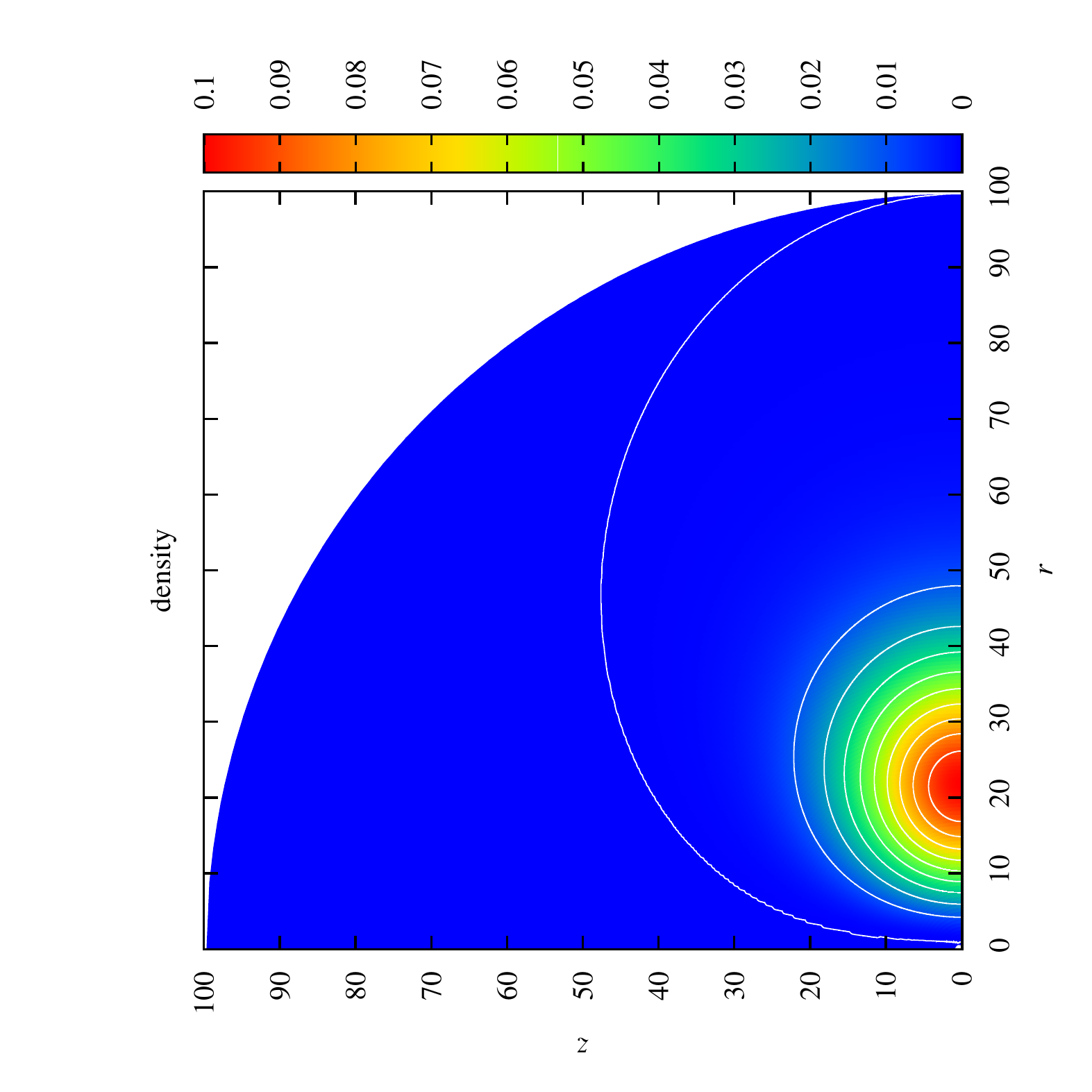}
\end{center}
\caption{An example of the density distribution in the selfgravitating rotating disk. The plot shows a cross section of the upper hemisphere in a meridian plane. Spatial dimensions are expressed in solar radii $R_{\astrosun}$. The density is color coded in $\mathrm{g \cdot cm^{-3}}$.}
\label{fig_disk_example}
\end{figure}

In the present Section we shall test the quality of the estimate (\ref{mass_ineq})  by comparing it with  appropriate  numerical solutions describing  selfgravitating rotating disks. The numerical method used here is quite standard \cite{eriguchi_muller_85, nishida_eriguchi_lanza_92, mach_malec_10}. Here we will only sketch the main idea.

Eqs.~(\ref{euler_eq}) can be formally integrated yielding
\begin{equation}
\label{integrated_euler_eq}
h + \Phi + \Phi_\mathrm{c} = \tilde C,
\end{equation}
where $\Phi_\mathrm{c}$ is given by (\ref{centrifugal_pot}) and $\Phi$ satisfies (\ref{poisson_eq}). The integration constant $\tilde C$ is important, and its value depends on the solution. Gravitational potential $\Phi$ can be expressed in terms of the Green function for the Laplace operator
\begin{equation}
\label{grav_pot_formula}
\Phi(\mathbf x) = -\frac{G M_\mathrm{c}}{|\mathbf x|} - G \int_\Omega d^3x^\prime \frac{\rho(\mathbf x^\prime)}{|\mathbf x - \mathbf x^\prime|},
\end{equation}
where $M_\mathrm{c}$ denotes the central mass. In the implementation of the numerical method the singularity of the Green function at $\mathbf x^\prime = \mathbf x$ causes a problem, that we  avoid in the standard way---by expanding the above integral in terms of Legendre polynomials \cite{eriguchi_muller_85}.

We assume that the domain $\Omega$ ranges from $r = r_\mathrm{in}$ to $r = r_\mathrm{out}$ in the equatorial plane, and the maximal density within the disk reaches a fixed value $\rho_\mathrm{max}$. The rotation law is $\omega(r) = v_0 / r$, and the equation of state has the form $p = K \rho^{1 + 1/n}$.

The choice of the initial parameters is to a large extent arbitrary. We fix  values of $r_\mathrm{in}$, $r_\mathrm{out}$, $n$, $\rho_\mathrm{max}$ and $M_\mathrm{c}$, whereas constants $\tilde C$, $v_0$ and $K$ are computed once the solution is known. For $v_0$ we have
\[ v_0^2 = \frac{\Phi(r_\mathrm{out}, z = 0) - \Phi(r_\mathrm{in}, z = 0)}{\ln(r_\mathrm{out} / r_\mathrm{in})}, \]
where we use the fact that the enthalpy $h$ vanishes on $\partial \Omega$. Values of $\tilde C$ and $K$ can be obtained form Eq.~(\ref{integrated_euler_eq}) taken at $(r_\mathrm{in},z=0)$ (or $(r_\mathrm{out},z=0)$) and the point where $\rho = \rho_\mathrm{max}$ respectively.

The structure of the disk is obtained by an iterative procedure. One starts by assuming an initial density distribution in the toroidal shape ranging from $r_\mathrm{in}$ to $r_\mathrm{max}$ with the a priori given maximum value  $\rho_\mathrm{max}$. Then  the gravitational potential is found from the formula (\ref{grav_pot_formula}). In the next step we compute constants $v_0$, $\tilde C$ and $K$ corresponding to the assumed density distribution. Finally Eq.~(\ref{integrated_euler_eq}) can be used to obtain a new approximate solution for $h$ and the corresponding distribution of $\rho$. If Eq.~(\ref{integrated_euler_eq}) gives a negative value of $h$ in some region of interest then we assume that $h = 0$ there. In this way a new shape of the disk is obtained.  This three-step procedure is iterated until a satisfactory convergence is reached.

The disadvantage of such approach is that a good spatial resolution is required in order to obtain accurate solutions, and computations of the gravitational potential $\Phi $ given by Eq.~(\ref{grav_pot_formula}) become time consuming.

An example of a disk solution obtained by the procedure described above is shown on Fig.~\ref{fig_disk_example}. The solution was obtained for $r_\mathrm{in} = 1 R_{\astrosun}$, $r_\mathrm{out} = 100 R_{\astrosun}$, $n = 3$, $\rho_\mathrm{max} = 0.1 \mathrm{g \cdot cm^{-3}}$ and $M_\mathrm{c} = 1 M_{\astrosun}$. The corresponding mass of the disk is equal to $M = 1.8 \cdot 10^{3} R_{\astrosun} $, and the bound given by (\ref{mass_ineq}) is $M > 3.8 \cdot 10^{2} R_{\astrosun}$. Here $M_{\astrosun}$ and $R_{\astrosun}$ denote the solar mass and radius respectively.

 \begin{table}[t!]
\caption{Disk masses and their estimates computed from (\ref{mass_ineq}). Configurations listed in the table were
computed for $n = 3$ and $M_\mathrm{c} = 1 M_{\astrosun}$.}
\begin{center}
\begin{tabular}{ccccc}
$r_\mathrm{in} \; [R_{\astrosun}]$ & $r_\mathrm{out} \; [R_{\astrosun}]$ & $\rho_\mathrm{max} \; [\mathrm{g \cdot
cm^{-3}}]$ & $M \; [M_{\astrosun}]$ & Mass estimate $[M_{\astrosun}]$ \\
\hline
5    & 10    & $0.1$     & $1.4$            & $5.1 \cdot 10^{-2}$ \\
1    & 10    & 1         & 25               & 2.7 \\
50   & 100   & $10^{-2}$ & $1.8 \cdot 10^2$ & $7.3$ \\
500  & 1000  & $10^{-5}$ & $1.8 \cdot 10^2$ & $7.3$ \\
2500 & 5000  & $10^{-6}$ & $2.2 \cdot 10^3$ & $92$\\
1    & 1000  & $10^{-4}$ & $1.5 \cdot 10^3$ & $3.8 \cdot 10^2$ \\
1    & 100   & $0.1$     & $1.8 \cdot 10^3$ & $3.8 \cdot 10^2$ \\
1    & 5000  & $10^{-5}$ & $1.8 \cdot 10^4$ & $5.5 \cdot 10^3$ \\ 
\end{tabular}
\end{center}
\label{table_mass}
\end{table}

Table \ref{table_mass} summarizes results obtained for a couple of numerical solutions and analytic estimates. The first three columns show values of the initial parameters: the inner radius $r_\mathrm{in}$, the outer radius $r_\mathrm{out}$ and the maximal mass density $\rho_\mathrm{max}$, respectively. The last two columns display the mass of a disk  and  its estimate derived from (\ref{mass_ineq}). Let us point out that the geometric data, $r_\mathrm{in}$ and $r_\mathrm{out}$, are needed only for the numerical calculation. 

The obtained inequality is never saturated. Relatively heavy disks, with the mass exceeding the central mass by 3--4 orders of magnitude, have masses close to that predicted by (\ref{mass_ineq}). It is clear that the accuracy of the functional-analytic bound increases with the mass of the disk, and thus with the increase of the selfgravity. The exact numerical mass and the predicted mass differ by a factor of three in the heavy end of masses, and by a less than two orders of magnitude for light  disks.

\section{Conclusions}

We derive, using certain functional inequalities, an analytic estimate for the mass of selfgravitating axially symmetric stationary configurations of polytropic fluids. It is  valid for polytropic indices $n \geq 3$, both for rotating stars and for accretion disks with centrifugal potentials satisfying the condition $\Delta \Phi_\mathrm{c} \leq 0$. This  class of potentials is quite general and   includes  several common types of rotation, with the rigid  $\omega = \mathrm{const}$ and the $v$-const rotations. 

The bound on the mass is given in terms of the polytropic index, maximal density and maximal temperature of the gas. The obtained expression  is strikingly similar  to the mass formula  of the  Lane--Emden stars.  One can use this result in order to obtain  the Jeans inequality for equilibrium ideal gas  disks and non-spherical stars for the polytropic index $n=3$.   We believe that this   is the first rigorous derivation of the Jeans mass and of the Jeans inequality for stationary systems. 

The accuracy of the analytic bound increases with the mass of the disk. The exact numerical mass and the predicted mass differ by a factor of three in the heavy end of disk masses, and by a less than two orders of magnitude for lighter  disks.

\section*{Acknowledgments}

The research was carried out with the supercomputer ``Deszno'' purchased thanks to the financial support of the European Regional Development Fund in the framework of the Polish Innovation Economy Operational Program (contract no. POIG. 02.01.00-12-023/08).

\end{document}